\documentclass[12pt,letterpaper]{article}
\usepackage{amsmath,amssymb,pgf,pgfarrows,pgfnodes,float,appendix, hyperref}
\usepackage{graphicx}
\usepackage{subfigure}

\usepackage[margin=0.9in]{geometry}
\usepackage{pgfplots}
\usepackage{amsmath}
\usepackage{tikz}

\title{{\rm\footnotesize \qquad \qquad \qquad \qquad \qquad \ \qquad \qquad \qquad \ \ \ \ \ \                  RUNHETC-2019-22 }\vskip.5in    Fermi/Pauli Duality in Arbitrary Dimension}
\author{Tom Banks\\
Department of Physics and NHETC\\
Rutgers University, Piscataway, NJ 08854\\
E-mail: \href{mailto:tibanks@ucsc.edu}{tibanks@ucsc.edu}}

\date{}
\begin{document}
\maketitle

\begin{abstract}
We propose an exact map from commuting lattice spin systems with gauge interactions to fermionic models in an arbitrary number of dimensions. 

\end{abstract}

\section{Introduction}

The literature on the Quantum Hall Effect introduced the notion of "flux attachment" to change the statistics of particles in two space dimensions.  The basic idea was simple.  In two dimensions a delta function magnetic flux is a point object.  If we have two point particles that carry both electric charge and magnetic flux, then they will experience an Aharonov-Bohm phase change when they are exchanged.  This can make bosons into anyons.  A lucid exposition of this idea can be found in the lecture notes of Preskill\cite{topqcpreskill}.  In\cite{QMMI} I suggested that this could be generalized to $d$ space dimensions by attaching $d -2$ branes to charged particles.  The resulting objects are not pointlike, but if the gauge group is restricted to be $Z_2$, then they can be particles transforming under the double cover of the rotation group, assuming that the $d-2$ brane carries zero energy.  The existence of spinless fermions in models that do not have (at least discrete) Lorentz invariance is easily understood by noting that we can always append an "internal" $SU(2)$ symmetry to our definition of the rotation group and convert the natural half integral spin Aharonov-Bohm fermions into singlets of the new group.   The original purpose of this paper was to carry out this program very explicitly for lattice models.   The basic idea is that all violation of local commutivity should be understood in terms of the AB effect.

In continuum quantum field theory, the Hilbert space in an arbitrarily small region is infinite dimensional.   Bosonization formulae are direct relations between fermion and boson fields at a point.  In a lattice theory, one cannot ignore the fact that the fermion number density operator at a point has spectrum $(0,1)$.  Thus, as in the Jordan-Wigner construction of fermions in one dimension, one can at best hope to have exact relations between fermionic lattice field theories and models of commuting Pauli matrices.  This is what I intended to construct.  In the interim between the conception of this idea and an attempt to implement it, some papers by Kapustin\cite{kapustin} and collaborators appeared, which carried out the program in space dimension $2$ and $3$.  The constructions are somewhat complex and this paper began as a simple attempt to understand them and extend them to general dimension.  Other references on previous attempts at higher dimensional bosonization can be found in \cite{highdbos}.  

Instead, I've found an alternative construction, which is quite straightforward in any number of dimensions (at least on hypercubic lattices).  It appears to be related to old ideas connecting Ising domain walls to fermions rather than to flux attachment.  Curiously, the central actor remains a topological $Z_2$ lattice gauge theory, the same one usually associated loosely with flux attachment.  However, despite some effort, I've been unable to find any $d - 2$ branes lurking in the formalism.  Instead, the $d - 1$ form gauge invariance of the topological model restricts the usual canonical conjugates of the $Z_2$ link variables to lie on closed or infinite domain walls in the dual lattice.  It's the anti-commutivity of these domain walls with the Wilson lines associated to $Z_2$ charged spin operators, which makes hard core bosons into fermions.  

In the $2 + 1$ dimensional theory of flux attachment, one marries a charged boson to a fluxon, and changes the spin by half integer values as one changes the statistics.
The mechanism proposed here does not change the spin.  Although the fermion operators we construct are not manifestly rotation invariant, they differ from scalar operators only by a pure gauge factor that is equal to one on the unique gauge invariant state of the pure gauge theory.  Of course, if we try to find a relativistic limit of a lattice fermion system the Nielsen-Ninomiya doubling theorem will automatically give us the right spin statistics connection, but we can also construct undoubled continuum fermions with a Galilean invariant dispersion relation.  In retrospect the idea that higher dimensional spin could emerge from an AB flux could not possibly have worked in a lattice theory.  The number operator on a site is completely gauge invariant and has eigenvalues $\pm 1$.  There is no apparent spin degeneracy.   However, one must be careful about boundary conditions in defining fermion operators, because, as we shall see, they involve infinite string and $d - 1$ brane operators in the $Z_2$ gauge theory. It's possible that subtleties at infinity could change this conclusion, but that would imply that there were also bosonic fields in the theory whose properties were dependent on boundary conditions.  We will discuss these issues, without resolving them, in the conclusions.

\section{Topological Hamiltonian $Z_2$ Lattice Gauge Theory in General Dimension}

The fundamental variable of a $Z_2$ lattice gauge theory is a link variable $U(C_1)$, which takes on values $\pm 1$.  The canonical conjugate is a shift operator $V(\tilde{C}_{d - 1})$ which flips the two states of the link Hilbert space.  We will assign this variable a position on the elementary $d - 1$ dimensional face of the dual lattice\footnote{We work with hypercubic lattices only, as a consequence of the author's inability to visualize any other kind of regular lattice in general dimension.} through which the link $C_1$ passes.  We have
\begin{equation} V(\tilde{C}_{d - 1}) U(C_1) = - U(C_1) V(\tilde{C }_{d - 1}) . \end{equation}  These variables commute with the copies of the same algebra on other link-dual $d -  1$ face pairs.   

We want our theory to be topological so we will take the Hamiltonian to be zero.  
Consider the continuum action \begin{equation} \int B_{d - 1} d A_1 , \end{equation} for a $1$ form gauge potential in $d + 1$ space-time dimensions.  In the gauge where time components of both gauge potentials vanish, the spatial components of the $d - 1$-form $B$ are the canonical conjugates of the spatial component of the $1$ form, and vice versa.  However, the action is also invariant under
\begin{equation} B_{d - 1} \rightarrow B_{d - 1} + d \Lambda_{d - 2}, \end{equation} if the gauge parameter $\Lambda_{d - 2}$ falls rapidly enough at infinity.  The constraint equation imposed by the time component of $B$ sets the spatial curl $dA_1 = 0$ in the absence of sources, while the constraint equation imposed by the time component of $A_1$ sets the spatial exterior derivative $d B_{d - 1}$ to zero in the absence of sources.  

The translation of these gauge invariances into the lattice say that only products of $U$ operators around closed 1 cycles ({\it i.e.} Wilson loops) and products of $V_{d - 1}$ around closed $d - 1$ cycles ('t Hooft surfaces) are gauge invariant in the pure gauge theory.   The constraints say that all of these operators are equal to $1$.  So the gauge invariant Hilbert space has only a single state, the conventional gauge invariant state in which all Wilson loops are equal to $1$.

Now we want to introduce hard core bosons, generators of the Pauli algebra,  $\sigma_{\pm} (P)$, charged under $Z_2$ at each point of the lattice.  
We want to construct gauge invariant operators that create charged states, starting from the state where $N(P) = 0$ for all $P$.  This will of course change the Gauss law constraint, and allow open Wilson lines between two points where $\sigma_+ = 0$, or between one such point and infinity. To implement this, we have to specify boundary conditions "at infinity".  In order to have a canonical conjugate for each link operator $U(l)$ we choose the dual lattice to include $d - 1$ faces pierced by the links on the boundary of the lattice where the Pauli operators sit.  We do not include dual faces that could close dual hypercubes on the boundary, so there is no Gauss law constraint on the points on the boundary of the lattice.   Thus, a Pauli operator $\sigma_{\pm} (P)$ multiplied by a semi-infinite Wilson line from $P$ to the boundary will create a gauge invariant $Z_2$ charged state.  

Charged fermion operators are constructed by choosing a direction $\hat{n}$ on the lattice\footnote{We will discuss below the question of other choices connecting the point $P$ to the boundary.} and writing 

\begin{equation} \psi^{\dagger} (P,\hat{n}) \equiv \sigma_+ (P) V(\tilde{P}_{\hat{n}}) U(P,\infty, \hat{n}) , \end{equation} 
where we've chosen the straight Wilson line in the positive $\hat{n}$ direction and
$\tilde{P}_{\hat{n}}$ is the dual $ d - 1$ plane in the directions perpendicular to $\hat{n}$ immediately in the positive $\hat{n}$ direction from $P$. See Figure 1 for the construction of the operator.
\begin{figure}
\begin{center}
\includegraphics[scale=0.5]{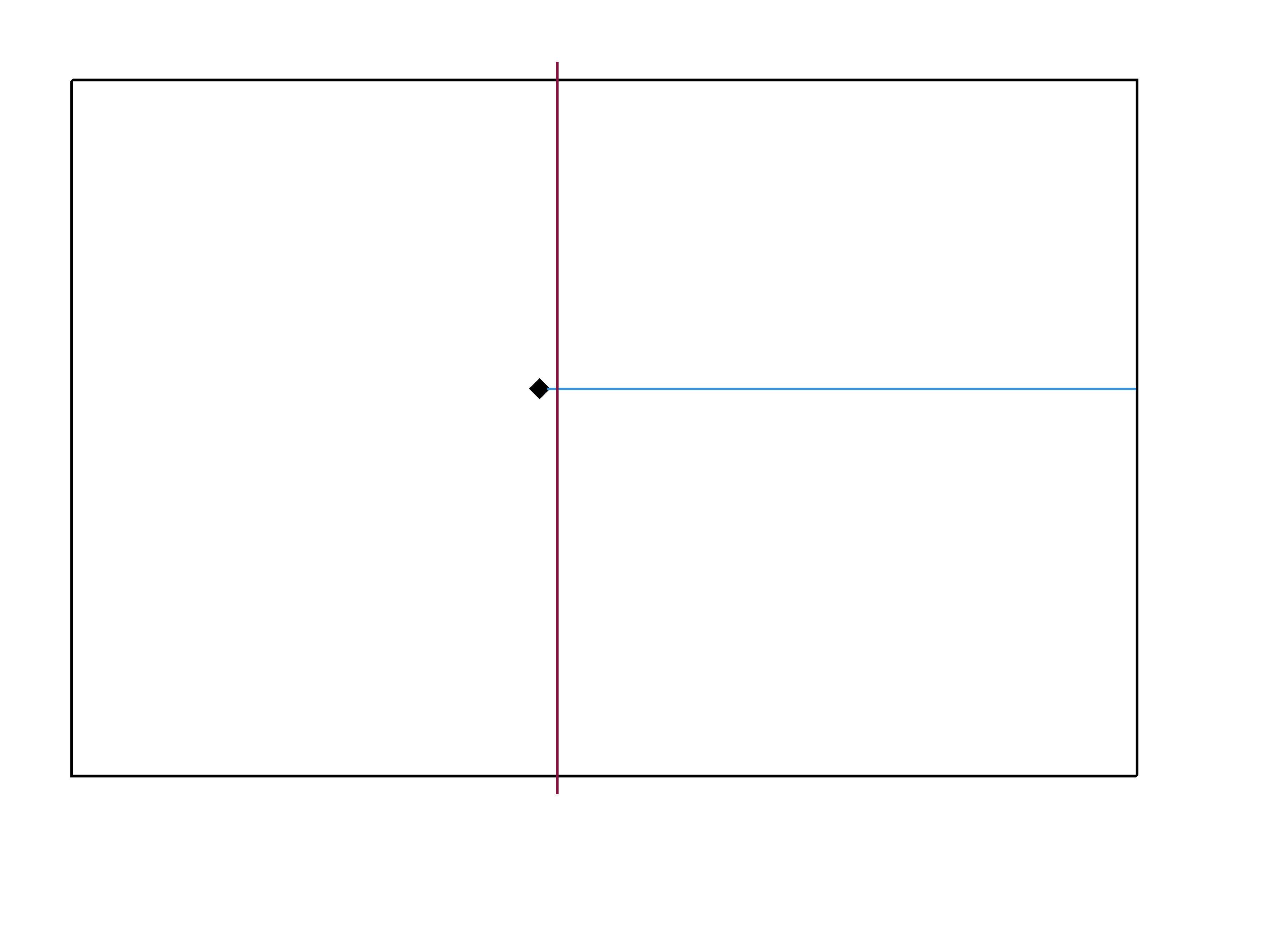}
\caption{The Construction of the Fermion Operator}
\label{k}
\end{center}
\end{figure}
Note that both the $U$ and $V$ operators are Hermitian because the gauge group is $Z_2$ .  The product of the fermion operator and its conjugate at two points $P,Q$ is  
\begin{equation}  V(\tilde{P}_{\hat{n}}) U(P,\infty, \hat{n})U(Q,\infty, \hat{n})V(\tilde{Q}_{\hat{n}})   \sigma_+ (P) \sigma_- (Q) .  \end{equation}  If $P = Q$ the gauge theory factor is equal to $1$ and the two operators anti-commute.   If $ P < Q$, where the notation means that the $\hat{n}$ component of $Q$ is larger than that of $P$, then the Wilson line from $P$ intersects the $d - 1$ plane $\tilde{P}_{\hat{n}}$ in exactly a single elementary $d - 1$ face, and the operators again anti-commute.

If we look at the product of $\psi (P)$ with $\psi (Q)$ then we have 

\begin{equation}  \psi (P)\psi (Q) = V(\tilde{P}_{\hat{n}}) U(P,\infty, \hat{n}) V(\tilde{Q}_{\hat{n}})  U(Q,\infty, \hat{n})V(\tilde{Q}_{\hat{n}})   \sigma_- (P) \sigma_- (Q) .  \end{equation} This vanishes if $P = Q$.  If $P < Q$ this is equal to 
\begin{equation}  \psi (P)\psi (Q) = - V(\tilde{P}_{\hat{n}}) V(\tilde{Q}_{\hat{n}}) U(P,\infty, \hat{n})  U(Q,\infty, \hat{n})V(\tilde{Q}_{\hat{n}})   \sigma_- (P) \sigma_- (Q) .  \end{equation}
\begin{equation} = - V(\tilde{P}_{\hat{n}}) V(\tilde{Q}_{\hat{n}}) U(P,Q, \hat{n})  \sigma_- (P) \sigma_- (Q) . \end{equation}  If we multiply in the opposite order, the minus sign does not appear because $U(P,\infty,\hat{n})$ does not penetrate the domain wall at $Q$.  The operators in the final expression are manifestly symmetric under interchange of $P$ and $Q$ because the gauge group is $Z_2$ so that $U(Q,P) = U^{-1} (P,Q) = U(P,Q)$, so the fermion operators anti-commute.

Let us define our Hilbert space by starting with the state $|0, 0 \rangle$ in the tensor product of the Pauli Hilbert space and that of the conventional $Z_2$ gauge theory. The notation means that we choose the state where $\sigma_+ (P) \sigma_- (P) $ equals zero for all $P$, tensored with the gauge invariant state in which all Wilson loops are equal to $1$.  Then we act on this state with all possible even functions of the fermion operators.    The matrix elements of all fermion bilinears between two states in this Hilbert space will be equal to the corresponding matrix elements of just the Pauli operators multiplied by a minus sign that depends on the initial and final states.   

This is, of course, exactly what we expect from standard fermion algebra.  Starting from the Fock vacuum of fermions $|0 \rangle_F$, we construct the Fock space as 
\begin{equation} \psi^{\dagger} (P_1 )\ldots \psi^{\dagger} (P_k) | 0 \rangle_F . \end{equation}
While it is certainly true that for a given choice of the $P_i$ there is only one state in the Hilbert space, completely characterized by the value of $N(P)$ (zero when $P$ is not one of the $P_i$ and one when it is), the value of $N(P)$ alone does not give us enough information to evaluate matrix elements of fermion bilinears in Fock space.
A particular ordering of the fermion creation operators can be thought of as a $Z_2$ Wilson line connecting the points $P_1 \ldots P_k$ in a certain order.  The matrix elements of fermion bilinears $\psi (P) \psi (Q)$ and $\psi^{\dagger} (Q)\psi (Q)$ for $P \neq Q$ are sensitive to certain aspects of the ordering.  They are not proportional to the corresponding bilinears of Pauli raising and lowering operators.  Thus, in order to construct fermion bilinears we require operators that do not commute with the Wilson lines.  The dual domain wall operators, which we introduced above, serve this purpose, though it is not clear if they are the most general construction.

Indeed, as we mentioned in the introduction, Kapustin and collaborators have introduced other constructions in two and three dimensions, which implement the notion of flux attachment and are equally consistent constructions of fermions out of spins and $Z_2$ gauge fields.  Even our domain wall construction of fermion operators is not unique.   We had to choose a direction $\hat{n}$ on the lattice as well as the sense of the Wilson line in the $\hat{n}$ direction.  That is, on a hypercubic lattice we can construct $2^d$ different Fermion fields $\psi_a (P)$ .  Under cubic rotations, which leave the point $P$ invariant, they transform among themselves.  The different components of $\psi_a (P)$ commute with each other.  They are not independent, since $\psi^{\dagger}_a \psi_a$ is equal to $N(P)$ for all $a$.

In fact, the different components of $\psi_a$ are gauge equivalent to each other in the Hilbert space we've defined.   To see this, note first that
\begin{equation} U(P,\hat{n},\infty) = U(P,\hat{m},\infty) U_{\hat{m},\hat{n}, P} , \end{equation} where the latter operator is a Wilson loop formed by connecting the ends of the two Wilson lines by a line on the boundary.   Furthermore, any closed Wilson loop commutes with the infinite domain wall operators because it penetrates a domain wall an even number of times.  Recall that we've used the natural definition of the dual lattice in which we include dual $d - 1$ faces that are pierced by the boundary links.(Fig. 2). 
\begin{figure}
\begin{center}
\includegraphics[scale=0.5]{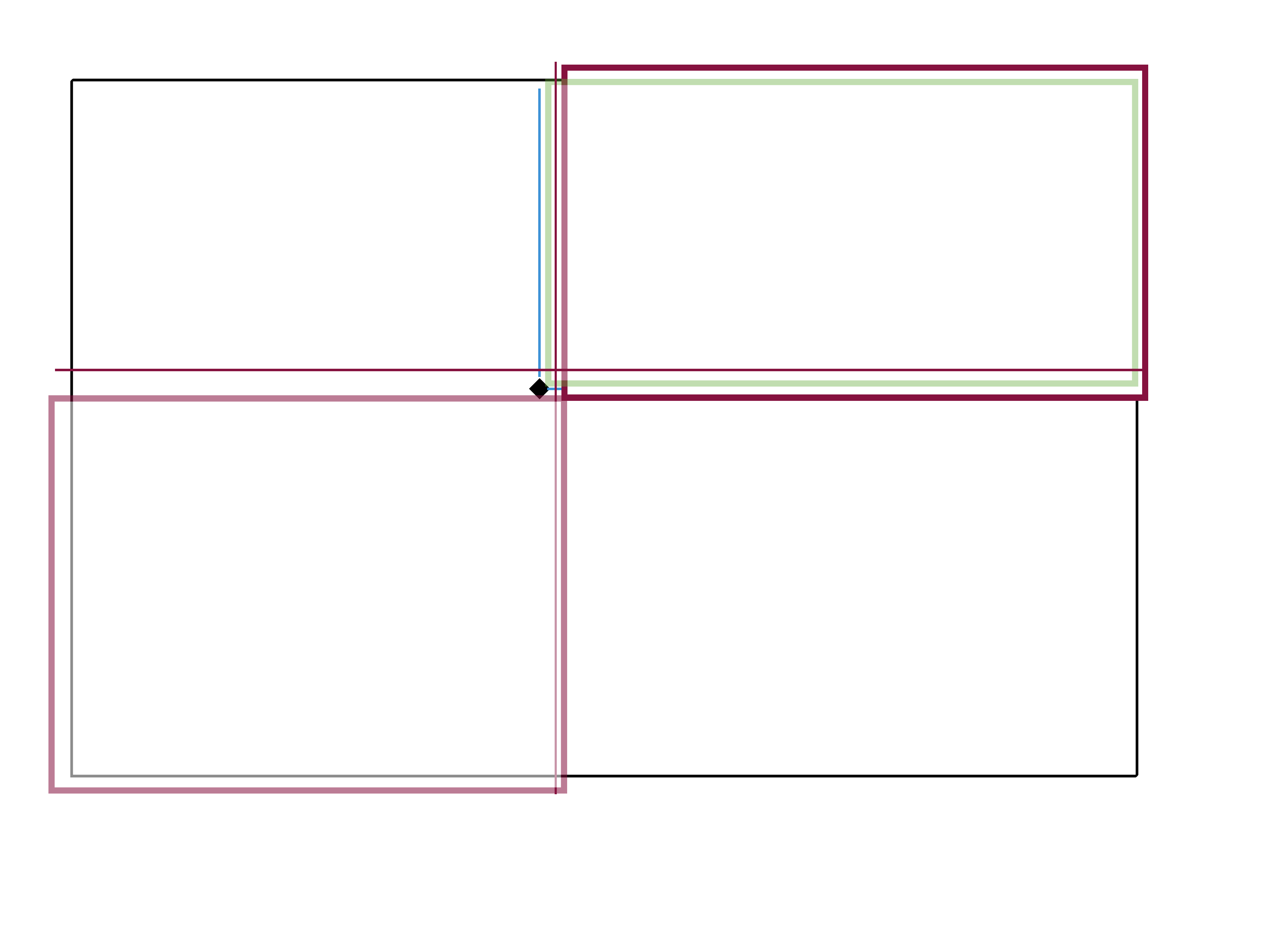}
\caption{Relation Between Different Choices of Axis For the Fermion Operator - Green Line is Wilson Loop, Magenta Closed d - 1 Surface}
\label{l}
\end{center}
\end{figure}
In a similar manner we can write the domain wall operator perpendicular to the $\hat{m}$ direction as the product of the domain wall operator perpendicular to the $\hat{n}$ and $V_{\hat{m},\hat{n}} (P) \equiv V(\tilde{P}_{\hat{m}})  V(\tilde{P}_{\hat{n}} )$. Thus
\begin{equation} \psi_{\hat{n}} (P) = \sigma_- (P) V (\tilde{P}_{\hat{m}}) V_{\hat{m},\hat{n}} (P) U(P,\hat{m},\infty) U_{\hat{m},\hat{n}} (P) . \end{equation} 

\begin{equation} \psi_{\hat{n}}^{\dagger} (P) = \sigma_+ (P)U(P,\hat{m},\infty) U_{\hat{m},\hat{n}} (P) V(\tilde{P}_{\hat{m}})   V_{\hat{m},\hat{n}}(P)  .\end{equation} 
Since the closed loop operators $U_{\hat{m}\hat{n}} (P)$ commute with all infinite domain walls and with each other, they can all be pushed to the right of any fermion operator to act on the unique state in the pure gauge theory and are all equal to $1$. 

To understand the action of the operators $V_{\hat{m},\hat{n}} (P_i)$ we use the fermion anti-commutation relations to write any even fermion operator, written in terms of $\psi_{\hat{n}} (P)$ with all the annihilation operators to the left and all the creation operators on the right.  Furthermore we order the operators according to their position along the $\hat{m}$ direction.  The largest $\hat{m}$ coordinate among annihilation operators is on the left and the largest among creation operators on the right.  
In this configuration we can move all of the $V_{\hat{m},\hat{n}} (P_i)$ operators so that they act on the pure gauge Hilbert space.  Since all of these operators commute with all Wilson loops and each of them squares to $1$, they take the state with all Wilson loops equal to $1$ into itself, multiplied by $\pm 1$. Furthermore they all give either a plus or a minus sign when commuted through an open Wilson line.   Thus, every product of even numbers of fermion creation and annihilation operators using the $\hat{n}$ convention is equal to a product of $\hat{m}$ operators at the same point, up to a reordering of anti-commuting operators.  The different definitions are all unitarily equivalent.  

 Thus, like the Dirac string of an abelian monopole, the apparent angular dependence of our fermion operators is a gauge artifact.  A similar argument works for any other uniform choice of the Wilson line that connects a point where a fermion acts to infinity.  The domain wall is parallel to the $d - 1$ face pierced by the first link of the Wilson line.  The commutation properties used above are purely topological.  They count the number of times modulo two that Wilson lines pierce domain walls.

\section{Discussion}

In the literature on the two dimensional Ising model, one reads that fermion operators are products of order times disorder operators.  In higher dimensions, disorder operators are associated with domain walls and this idea has been used to write a fermionic representation of the $2 + 1$ dimensional transverse Ising/$Z_2$ lattice gauge theory\cite{fradkinetal} .   The fermi/pauli dictionary in this paper uses zero energy domain walls of a topological $Z_2$ gauge theory with a one dimensional Hilbert space.   Using that dictionary, any fermion Hamiltonian can be written in terms of spins and any gauge invariant spin Hamiltonian can be written in terms of fermions, in any number of dimensions.  The essential fact that makes this work is that the matrix elements of even numbers of fermion fields between states in fermionic Fock space differ from those of commuting Pauli matrices only by minus signs.   The topological gauge theory gives us a machine for keeping track of those minus signs.   It remains to be seen whether this insight can be used to solve previously unsolved models or to aid in computational approaches to fermion problems.   One possibility that comes to mind, in a functional integral version of this mapping, is to somehow "do the boson dynamics for fixed values of the lattice gauge fields", using {\it e.g.} Monte Carlo methods and isolate the sign problem in the final gauge field integration.  

It also seems important to understand the relation between the formulae in this paper and the work on lattice flux attachment\cite{kapustin}, which, at least in some cases is an alternate way to relate fermions to spins.  Our construction does not map local Hamiltonians even in fermi fields to local Hamiltonians of spins.  However, the non-locality is entirely in the pure gauge sector and the pure gauge Hilbert space has only a single state and the non-local operators simply multiply that state by an appropriate minus sign.  The domain walls and Wilson lines in our construction carry no energy.  Gaiotto and Kapustin\cite{gk} have given a general argument that local bosonization requires one to have an {\it anomalous} $d - 2$ form gauge symmetry ($d - 2$ is the rank of the gauge parameter), whereas our formalism has a gauged $d - 2$ form symmetry.  The lattice models of \cite{kapustin} implement the observation of\cite{gk} .
Since breaking of a gauge symmetry can always be attributed to the Brout-Englert-Higgs mechanism, it might be possible to obtain those models by adding another $Z_2$ valued $d -2$ face variable to our setup. This would introduce the possibility of constructing $d - 2$ branes, and also of obeying the Gaiotto-Kapustin criterion.

In this connection, we should point out a fundamental problem with the notion of flux attachment on a lattice that preserves a non-abelian subgroup of the rotation group.  The arguments for flux attachment in continuum single particle mechanics in two dimensions, show that the spin is shifted by $1/2$ when changing bosons to fermions.  In a hypothetical $d - 2$ brane attachment procedure in $d$ space dimensions, the transverse dimensions to the $d - 2$ brane can lie in any plane and the $U(1))$ generator of rotations in that plane gets its eigenvalues shifted by a half integer.  On a lattice there will of course be special planes but if there is a non-abelian subgroup of rotations preserved then the spinor representation of the rotation subgroup will have dimension $> 1$.   If we start from a system that has only a single Pauli algebra on each site, then the space of gauge invariant states at a point is two dimensional and cannot represent a full spinor's worth of fermion operators.  The constructions of \cite{kapustin} avoid this contradiction because their fermion construction explicitly breaks the planar rotation symmetries of the lattice.  They also give only a single spin component per lattice site.  Thus it appears that the naive continuum ideas of flux attachment cannot be implemented in lattice models.

For systems whose low energy dispersion relation is that of a massless relativistic fermion, the Nielsen-Ninomiya theorem\cite{nn} provides an elegant way out of this problem of fermion spin.   The topological NN doubling phenomenon assures us that a single component lattice fermion will converge to a multiple component continuum fermion field, built from lattice fermions at the vertices of a unit cell\footnote{Unfortunately, NN doubling is overkill.  It does not allow us to construct chiral fermions, and usually gives us more Dirac fermions than we would like to have in order to reproduce a general continuum gauge theory. As a consequence there can be relevant operators in the continuum limit, which break rotation invariance and preserve only a diagonal subgroup of spin and flavor rotations. This does not occur in lattice models consisting of only gauge fields and fermions.}.  Our Fermi/Pauli duality is independent of the lattice fermion Hamiltonian and so can be applied to systems with emergent Lorentz invariance.  It can produce relativistic systems that satisfy the spin-statistics theorem, once one has dealt with possible relevant operators that violate rotation invariance..  On the other hand, if the fermion Hamiltonian has emergent Galilean symmetry, then one lattice Fermion field produces one continuum Fermion field and our duality will produce spinless Fermions unless we append spin as an internal symmetry.  Of course, such spinless Fermions abound in condensed matter models, if not necessarily in the real world.  

It's possible that we could exploit the sensitivity to boundary conditions of our Fermion construction to append spin to the Fermion operators by adding spin multiplets on the boundary.  Indeed, the famous demonstrations involving belts or hand held water glasses\cite{belts}, which reveal the $SU(2)$ covering group of $SO(3)$, suggest that the rotational properties of spinors do somehow depend on behavior at infinity.  However, a hypothetical construction of this sort would imply that the integer spin local composite operators built out of Fermions were also sensitive to boundary conditions at infinity, and this seems less palatable.  

That being said, we should mention one subtlety of our construction, relating to boundary conditions at infinity.  Let's choose the definition of fermions with Wilson lines in the positive $\hat{n}$ direction.  Consider the points on the $d - 1$ face of the boundary perpendicular to $\hat{n}$.  There are Pauli operators on those points, but no Wilson lines pointing in the positive $\hat{n}$ direction and no dual $d - 1$ face parallel to the boundary in the positive $\hat{n}$ direction.  Thus, the excitations on that boundary face are bosons rather than fermions.  To repair this we would have to add some boundary degrees of freedom, but this would change our discussion of the equivalence between different choices of the Wilson line that connects charges to the boundary.  This indicates that our definition of fermions {\it is} sensitive to the topology of manifolds on which they're defined.  

Finally, let me remark that the paper \cite{moorefreed} pointed out the generic existence of fermion operators in continuum models of self dual p-form gauge fields.  It's likely that the constructions in this note are related to that observation, although I have not yet found the connection.

\begin{center} 
{\bf Acknowledgments }  \end{center}

I would like to thank G.Moore, N.Seiberg, and especially A. Kapustin for email conversations about flux attachment and for reading a draft of this manuscript before publication. The work of T.Banks is partially supported by the U.S. Department of Energy under Grant DE-SC0010008. This research was supported in part by the National Science Foundation under Grant No. NSF PHY-1748958 because his paper was completed while the author was in residence at the KITP in Santa Barbara.

\end{document}